\documentclass[conference]{IEEEtran}
\usepackage[dvips]{graphicx}
\usepackage[cmex10]{amsmath}
\usepackage{multirow}
\usepackage{epsfig}
\usepackage{mathrsfs}
\usepackage{amssymb}
\usepackage{comment}
\usepackage{bm}
\usepackage{array}
\usepackage{amsthm}
\usepackage{blkarray}
\usepackage{fancyhdr}
\usepackage{enumerate}

\renewcommand{\thispagestyle}[1]{} 

\theoremstyle{definition}
\newtheorem{theorem}{Theorem}[section]
\newtheorem{lemma}[theorem]{Lemma}

\newtheorem{proposition}[theorem]{Proposition}

\ifCLASSINFOpdf

\else

\fi

\usepackage[tight,footnotesize]{subfigure}

\begin{document}
\pagestyle{fancy}
\IEEEoverridecommandlockouts

\lhead{\textit{Technical Report, Dept. of EEE, Imperial College, London, UK, May, 2012.}}
\rhead{} 
%
\title{Additive Link Metrics Identification: Proof of Selected Lemmas and Propositions}
\author{\IEEEauthorblockN{Liang Ma\IEEEauthorrefmark{2}, Ting He\IEEEauthorrefmark{3}, Kin K. Leung\IEEEauthorrefmark{2}, Don Towsley\IEEEauthorrefmark{1}, and Ananthram Swami\IEEEauthorrefmark{4}}
\IEEEauthorblockA{\IEEEauthorrefmark{2}Imperial College, London, UK. Email: \{l.ma10, kin.leung\}@imperial.ac.uk\\
\IEEEauthorrefmark{3}IBM T. J. Watson Research Center, Hawthorne, NY, USA. Email: the@us.ibm.com\\
\IEEEauthorrefmark{1}University of Massachusetts, Amherst, MA, USA. Email: towsley@cs.umass.edu\\
\IEEEauthorrefmark{4}Army Research Laboratory, Adelphi, MD, USA. Email: ananthram.swami.civ@mail.mil
}
}

\maketitle

\IEEEpeerreviewmaketitle

\section{Introduction}
Selected lemmas and propositions in \cite{MaNetworkTomography12} are proved in detail in this report. We first list the lemmas and propositions in Section II and then give the corresponding proofs in Section III. See the original paper \cite{MaNetworkTomography12} for terms and definitions.

\section{Lemmas and Propositions}
Let $\mathcal{H}$ denote the interior graph of graph $\mathcal{G}$, where two monitoring nodes ($m_1$ and $m_2$) are employed. In this report, Conditions \textcircled{\small 1} \normalsize and \textcircled{\small 2} \normalsize refer to the two following conditions.
\begin{description}
  \item[\textcircled{\small 1}]\normalsize $\mathcal{G}-l$ is 2-edge-connected for each interior link $l$ in $\mathcal{H}$;\looseness=-1
  \item[\textcircled{\small 2}]\normalsize $\mathcal{G}+m_1m_2$ is 3-vertex-connected.
\end{description}

\begin{lemma}
\label{Lemma:MeasurementMatrixReconstruction}
When the interior graph $\mathcal{H}$ of $\mathcal{G}$ is connected, the corresponding measurement matrix $\mathbf{R}$ can be linearly transformed to
\scriptsize
\begin{equation}\label{matrix3-2}
\left. \begin{aligned}
\left.
        \begin{blockarray}{ccccccccccc}
            \BAmulticolumn{5}{c}{\multirow{1}{*}{$W_{m_1a_1}\cdots W_{m_1a_{k_1}}$}}&\BAmulticolumn{4}{c}{\multirow{1}{*}{$W_{b_1m_2}\cdots W_{b_{k_2}m_2}$}}&\BAmulticolumn{2}{c}{\multirow{1}{*}{$W_{l_1}\cdots W_{l_{k_h}}$ }}\\
            \begin{block}{(ccccc|cccc|cc)}
         1 & && & & 1 &   &  &  &  \BAmulticolumn{2}{c}{\multirow{5}{*}{\Large $\mathbf{B}$}}  \\
         1 & && & &  & 1  &  &  &    &    \\
         \vdots &  & & & & &   & \ddots &  &    &  \\
         1 & & & &   & &   &  & 1 &    &    \\
         \cline{1-11}
         -1&1&  & & &   &   &  &  &   \BAmulticolumn{2}{c}{\multirow{5}{*}{\Large $\mathbf{T}$}}     \\
         -1&& 1 & & &  &    &  &  &    &  \\
         \vdots&& &\ddots & & &   &   &  &    &   \\
         -1&&  & & 1  & &   &  &   &    &   \\
         \cline{1-11}
         &  & &   &  &  & & & & \BAmulticolumn{2}{c}{\multirow{4}{*}{\Large $\mathbf{L}$}}\\
         &  &&    &  &  &  &&&  &            \\
        &  &&    &  &  &  &&&  &           \\
        & & &    &  &  &  &&&  &            \\
    \end{block}
  \end{blockarray}\right.\\
  \end{aligned} \right.,
\end{equation}
\normalsize
where the rest entries are zero, $\mathbf{B}$ is a $k_2\times k_h$ Boolean matrix\footnote{\emph{Boolean matrix} is a matrix each of whose entries is 0 or 1.}, $\mathbf{T}$ is a ($-1,0,1$)-Matrix\footnote{\emph{($-1,0,1$)-Matrix} is a matrix each of whose entries is $-1$, 0, or 1.} with dimensions $(k_1-1)\times k_h$, and $\mathbf{L}$ with $k_h$ columns is a matrix associated with all rows in the restructured measurement matrix $\mathbf{R}$ involving only $(W_{l_i})^{k_h}_{i=1}$.
\end{lemma}

\begin{proposition}
\label{lemma:3vertexConnected}
For graph $\mathcal{G}$, if all link metrics in the associated interior graph are identifiable through simple path measurements, then $\mathcal{G}+m_1m_2$ is 3-vertex-connected.
\end{proposition}

\begin{proposition}
\label{Lemma-3-connected}
Using two monitoring nodes, the necessary and sufficient condition for $\mathcal{G}+m_1m_2$ being a 3-vertex-connected graph is when 2 nodes are deleted in $\mathcal{G}$, the remaining graph is still connected, \emph{or} every connected component each has a monitoring node.
\end{proposition}

\begin{lemma}
\label{Lemma-twoCycles}
If graph $\mathcal{G}$ satisfies Conditions \textcircled{\small 1} \normalsize and \textcircled{\small 2}\normalsize ,
for any link $vw$ in the interior graph of $\mathcal{G}$, two cycles ($\mathcal{C}_1$ and $\mathcal{C}_2$) can be discovered in $\mathcal{G}$, such that
\begin{enumerate}[(a)]
  \item $\mathcal{C}_1$ is a face\footnote{In the sequel, $\mathcal{C}_1$ means a cycle as well as a face.};
  \item $vw$ is the only common link between $\mathcal{C}_1$ and $\mathcal{C}_2$;
  \item $\mathcal{C}_1$ and $\mathcal{C}_2$ have \emph{one} common node at most, apart from $v$ and $w$;
  \item there exists path $\mathcal{P}_1$ connecting\footnote{Let $m^*_1, m^*_2 \in \{m_1,m_2\}$ with $m^*_1\neq m^*_2$.} $m^*_1$ and a node on $\mathcal{C}_1-v-w$ and $\mathcal{P}_2$ connecting $m^*_2$ and a node on $\mathcal{C}_2-v-w$;
  \item $\mathcal{P}_1 \cap \mathcal{P}_2=\emptyset$;
  \item $L(\mathcal{P}_1)\cap L(\mathcal{C}_1-v-w)=\emptyset$, $L(\mathcal{P}_2)\cap L(\mathcal{C}_2-v-w)=\emptyset$;
  \item $v,w\notin V(\mathcal{P}_1)$ and $v,w\notin V(\mathcal{P}_2)$.
\end{enumerate}
\end{lemma}

\begin{proposition}
\label{Proposition:FaceOneBorderLink}
If graph $\mathcal{G}$ satisfies Conditions \textcircled{\small 1} \normalsize and \textcircled{\small 2}\normalsize , then
\begin{enumerate}[(a)]
    \item for any face in graph $\mathcal{G}$, there is at most \emph{one} border-link in this face;
    \item for any border-link $vw$ in the interior graph of $\mathcal{G}$, it can discover a face without traversing $m_1$ and $m_2$;
    \item for every border-link $vw$ on face $\mathcal{C}_1$ with $L(\mathcal{C}_1)\subseteq L(\mathcal{H})$, there exist paths $\mathcal{P}(m_1,v)$ and $\mathcal{P}(m_2,w)$ with $\mathcal{P}(m_1,v) \cap \mathcal{P}(m_2,w)=\emptyset$, $\mathcal{P}(m_1,v)\overset{\circ}{v} \cap \mathcal{C}_1=\emptyset$ and $\mathcal{P}(m_2,w)\overset{\circ}{w} \cap \mathcal{C}_1=\emptyset$.
\end{enumerate}
\end{proposition}

\section{Proofs}

\subsection{Proof of Lemma \ref{Lemma:MeasurementMatrixReconstruction}}
\begin{figure}[tb]
\vspace{-.5em}
\centering
\includegraphics[width=2.2in]{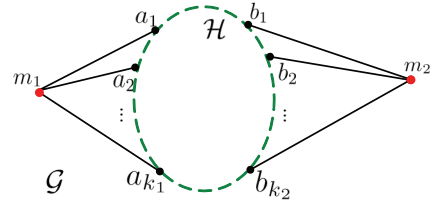}
\caption{Reorganized graph $\mathcal{G}$, where $a_i$ and $b_j$ can be the same node.} \label{ReorganizedGraph}
\vspace{-6mm}
\end{figure}
Suppose that the two monitoring nodes $m_1$ and $m_2$ give rise to a connected interior graph $\mathcal{H}$ in $\mathcal{G}$ (Fig. \ref{ReorganizedGraph}). Let $(l_i)^{k_h}_{i=1}:=L(\mathcal{H})$ be the link set of $\mathcal{H}$ with $k_h:=||\mathcal{H}||$. Consider the measurement matrix $\mathbf{R_{a_i}}$ ($i\in \{1,\cdots,k_1\}$) corresponding to all possible paths $m_1\to a_i\to\ldots\to b_j\to m_2$ ($j=1,\ldots,k_2$). Then $\mathbf{R_{a_i}}$ is of the form
\scriptsize
\begin{center}
$
\left. \begin{aligned}
\mathbf{R_{a_i}}=&\left.
        \begin{blockarray}{ccccccc}
            W_{m_1a_i}&\BAmulticolumn{4}{c}{\multirow{1}{*}{$W_{b_1m_2}\cdots W_{b_{k_2}m_2}$}}&\BAmulticolumn{2}{c}{\multirow{1}{*}{$W_{l_1}\cdots W_{l_{k_h}}$ }}\\
            \begin{block}{(c|cccc|cc)}
         1 & 1 &   &  &  &   \BAmulticolumn{2}{c}{\multirow{4}{*}{\Large $\mathbf{B_{i1}}$}}    \\
         \vdots  & \vdots &   &  &  &    &      \\
         1   & 1 &   &  &  &   &       \\
         \cline{1-7}
         1   &  & 1  &  &  &   \BAmulticolumn{2}{c}{\multirow{4}{*}{\Large $\mathbf{B_{i2}}$}}      \\
         \vdots  &  & \vdots  &  &  &    &     \\
         1   &  & 1  &  &  &   &     \\
         \cline{1-7}
         \vdots    &  &    & \ddots &  &    \BAmulticolumn{2}{c}{{\vdots}}      \\
         \cline{1-7}
         1    &  & &  & 1 &    \BAmulticolumn{2}{c}{\multirow{4}{*}{\Large $\mathbf{B_{ik_2}}$}}      \\
         \vdots    &  & &  & \vdots &   &     \\
         1    &  & &  & 1 &    &      \\
    \end{block}
  \end{blockarray}\right.\\
  \end{aligned} \right.,
$
\end{center}
\normalsize
where $\mathbf{B_{ij}}$ ($j=1,\ldots,k_2$) with $k_h$ columns is the Boolean matrix corresponding to all simple paths between $a_i$ and $b_j$ in $\mathcal{H}$ (i.e., the $(p,\: q)$-th entry indicates whether link $l_q$ in $\mathcal{H}$ appears on the $p$th path between $a_i$ and $b_j$). Note that at least one such path exists, i.e., $\mathbf{B_{ij}}$ is nonempty, since $\mathcal{H}$ is connected. Among all rows containing $W_{m_1a_i}$  and $W_{b_jm_2}$ in $\mathbf{R_{a_i}}$, if we subtract the first row from the others, then the other rows are only non-zero in entries corresponding to $(W_{l_i})^{k_h}_{i=1}$. Reorganizing these subtracted rows to the bottom, we transform $\mathbf{R_{a_i}}$ to
\scriptsize
\begin{equation}\label{matrix2-2}
\left. \begin{aligned}
&\mathbf{R'_{a_i}}=\left.
        \begin{blockarray}{ccccccc}
            W_{m_1a_i}&\BAmulticolumn{4}{c}{\multirow{1}{*}{$W_{b_1m_2}\cdots W_{b_{k_2}m_2}$}}&\BAmulticolumn{2}{c}{\multirow{1}{*}{$W_{l_1}\cdots W_{l_{k_h}}$ }}\\
            \begin{block}{(c|cccc|cc)}
         1 & 1 &   &  &  &   \BAmulticolumn{2}{c}{\mathbf{r_{i1}}}   \\
         1 &   &  1 &  &  &  \BAmulticolumn{2}{c}{\mathbf{r_{i2}}}   \\
         \vdots  &  &   & \ddots &  &   \BAmulticolumn{2}{c}{    \vdots}    \\
         1 &   &    &  &  1 &   \BAmulticolumn{2}{c}{ \mathbf{r_{ik_2}}}    \\
         \cline{1-7}
             &   &   &  &  &  \BAmulticolumn{2}{c}{\multirow{3}{*}{\Large $\mathbf{L_{a_i}}$}}\\
             &  &    &  &  &    &            \\
             &  &    &  &  &    &            \\
    \end{block}
  \end{blockarray}\right.\\
  \end{aligned} \right.,
  \vspace{-1.6em}
\end{equation}
\normalsize
where $\mathbf{r_{ij}}$ ($j=1,\ldots,k_2$) is the first row of $\mathbf{B_{ij}}$, and $\mathbf{L_{a_i}}$ is a matrix derived from the subtraction operation. Combining all $\mathbf{R'_{a_i}}$ \big($i\in \{1,\cdots,k_1\}$\big) in (\ref{matrix2-2}), the restructured measurement matrix $\mathbf{R}$ is
\scriptsize
\begin{equation}\label{matrix3}
\left. \begin{aligned}
&\left.
       \begin{blockarray}{cccccccccc}
\BAmulticolumn{4}{c}{\multirow{1}{*}{$W_{m_1a_1}\cdots W_{m_1a_{k_1}}$}}&\BAmulticolumn{4}{c}{\multirow{1}{*}{$W_{b_1m_2}\cdots W_{b_{k_2}m_2}$}}&\BAmulticolumn{2}{c}{\multirow{1}{*}{$W_{l_1}\cdots W_{l_{k_h}}$ }}\\
            \begin{block}{(cccc|cccc|cc)}
         1 & & & & 1 &  &   &  &    \BAmulticolumn{2}{c}{ \mathbf{r_{11}}}     \\
         1 & & & &  & 1 &   &  &    \BAmulticolumn{2}{c}{ \mathbf{r_{12}} }   \\
         \vdots   & & & &&    & \ddots &  &    \BAmulticolumn{2}{c}{ \vdots }  \\
         1  & & &   & & &    & 1 &    \BAmulticolumn{2}{c}{\mathbf{r_{1k_2}}}    \\
         \cline{1-10}
         &1 & & & 1 & &    &  &    \BAmulticolumn{2}{c}{ \mathbf{r_{21}} }  \\
         &1 & & & &1 &    &  &   \BAmulticolumn{2}{c}{ \mathbf{r_{22}} }   \\
         & \vdots & & && &    \ddots &  &   \BAmulticolumn{2}{c}{ \vdots }  \\
         &1 & &   & &  &   & 1 &   \BAmulticolumn{2}{c}{ \mathbf{r_{2k_2}} }   \\
         \cline{1-10}
         & &\ddots &  & &  &     \vdots &  &     \BAmulticolumn{2}{c}{ \vdots  }   \\
         \cline{1-10}
         & & &1 & 1 & &  &    &    \BAmulticolumn{2}{c}{ \mathbf{r_{k_11}}  }  \\
         & & &1 & &1 &  &    &    \BAmulticolumn{2}{c}{ \mathbf{r_{k_12}}  }  \\
         & & &\vdots && &    \ddots &  &    \BAmulticolumn{2}{c}{ \vdots  }  \\
         & & &1   & &  &   & 1 &   \BAmulticolumn{2}{c}{ \mathbf{r_{k_1k_2}} }  \\
         \cline{1-10}
         &   &   &  & &  & & & \BAmulticolumn{2}{c}{\multirow{3}{*}{\Large $\mathbf{L_1}$}}\\
         &  &    &  & &   &&&  &           \\
        &  &    &  & &   &&&  &            \\
    \end{block}
  \end{blockarray}\right.
  \end{aligned} \right.,
\end{equation}
\normalsize
where $\mathbf{L_1}$ is the matrix formed by arranging $\mathbf{L_{a_1}},\ldots,\mathbf{L_{a_{k_1}}}$ vertically.
We apply the following linear transformations to (\ref{matrix3}): (i) first, subtracting row $i$ from row $qk_2+i$ for each $q=1,\ldots,k_1-1$ and $i=1,\ldots,k_2$; (ii) then, subtracting row $qk_2+1$ from row $qk_2+i$ for each $q=1,\ldots,k_1-1$ and $i=2,\ldots,k_2$; (iii) finally, moving all rows containing $r_{ij}-r_{1j}-r_{i1}$ ($i=2,\cdots,k_1$ and $j=2,\cdots,k_2$) to the matrix bottom. Ignoring entries whose values are zeros ($(W_{m_1a_i})^{k_1}_{i=1}$ and $(W_{b_jm_2})^{k_2}_{j=1}$ are not involved in the rows containing $r_{ij}-r_{1j}-r_{i1}$ ($i=2,\cdots,k_1$ and $j=2,\cdots,k_2$)), (\ref{matrix3}) is transformed into

\scriptsize
\begin{equation}\label{matrix3-2}
\left. \begin{aligned}
&\left.
        \begin{blockarray}{ccccccccccc}
            \BAmulticolumn{5}{c}{\multirow{1}{*}{$W_{m_1a_1}\cdots W_{m_1a_{k_1}}$}}&\BAmulticolumn{4}{c}{\multirow{1}{*}{$W_{b_1m_2}\cdots W_{b_{k_2}m_2}$}}&\BAmulticolumn{2}{c}{\multirow{1}{*}{$W_{l_1}\cdots W_{l_{k_h}}$ }}\\
            \begin{block}{(ccccc|cccc|cc)}
         1 & && & & 1 &   &  &  &    \BAmulticolumn{2}{c}{\mathbf{r_{11}}}   \\
         1 & && & &  & 1  &  &  &   \BAmulticolumn{2}{c}{ \mathbf{r_{12}}  }  \\
         \vdots &  & & & & &   & \ddots &  &   \BAmulticolumn{2}{c}{ \vdots  }  \\
         1 & & & &   & &   &  & 1 &   \BAmulticolumn{2}{c}{  \mathbf{r_{1k_2}}}    \\
         \cline{1-11}
         -1&1&  & & &   &   &  &  &    \BAmulticolumn{2}{c}{ \mathbf{r_{21}}-\mathbf{r_{11}} }   \\
         -1&& 1 & & &  &    &  &  &    \BAmulticolumn{2}{c}{ \mathbf{r_{31}}-\mathbf{r_{11}}}     \\
         \vdots&& &\ddots & & &   &   &  &    \BAmulticolumn{2}{c}{ \vdots }   \\
         -1&&  & & 1  & &   &  &   &   \BAmulticolumn{2}{c}{ \mathbf{r_{k_11}}-\mathbf{r_{11}} }  \\
         \cline{1-11}
         &  & &   &  &  & & & & \BAmulticolumn{2}{c}{\multirow{4}{*}{\Large $\mathbf{L}$}}\\
         &  &&    &  &  &  &&&  &            \\
        &  &&    &  &  &  &&&  &            \\
        & & &    &  &  &  &&&  &            \\
    \end{block}
  \end{blockarray}\right.\\
  \end{aligned} \right.,
\end{equation}
\normalsize
where
\begin{center}
$\mathbf{L}:=$
\scriptsize
$
\left. \begin{aligned}
&\left.
        \begin{blockarray}{cc}
            \BAmulticolumn{2}{c}{\multirow{1}{*}{$W_{l_1}\ \ W_{l_2} \cdots\ \  W_{l_{k_h}}$ }}\\
            \begin{block}{(cc)}
        \BAmulticolumn{2}{c}{\multirow{2}{*}{\Large $\mathbf{L_1}$}}\\
        &            \\
        \BAmulticolumn{2}{c}{ \mathbf{r_{22}-r_{12}-r_{21}} }\\
        \BAmulticolumn{2}{c}{ \mathbf{r_{23}-r_{13}-r_{21}} }\\
        \BAmulticolumn{2}{c}{ \vdots }            \\
        \BAmulticolumn{2}{c}{ \mathbf{r_{2k_2}-r_{1k_2}-r_{21}} }\\
        \BAmulticolumn{2}{c}{ \mathbf{r_{32}-r_{12}-r_{31} }}\\
        \BAmulticolumn{2}{c}{ \mathbf{r_{33}-r_{13}-r_{31} }}\\
        \BAmulticolumn{2}{c}{ \vdots }            \\
        \BAmulticolumn{2}{c}{ \mathbf{r_{3k_2}-r_{1k_2}-r_{31}} }\\
        \BAmulticolumn{2}{c}{ \vdots }            \\
        \BAmulticolumn{2}{c}{ \mathbf{r_{k_12}-r_{12}-r_{k_11}} }\\
        \BAmulticolumn{2}{c}{ \mathbf{r_{k_13}-r_{13}-r_{k_11}} }\\
        \BAmulticolumn{2}{c}{ \vdots }            \\
        \BAmulticolumn{2}{c}{ \mathbf{r_{k_1k_2}-r_{1k_2}-r_{k_11}} }\\
    \end{block}
  \end{blockarray}\right.\\
  \end{aligned} \right..
$
\end{center}

Entries in $\mathbf{r_{i1}-r_{11}}$ are of the value of -1, 0, or 1, since each entry in $\mathbf{r_{ij}}$ is 0 or 1. Therefore,
$\mathbf{B}:=$
\scriptsize
$
\left. \begin{aligned}
&\left.
        \begin{blockarray}{cc}
            \begin{block}{(cc)}
        \BAmulticolumn{2}{c}{ \mathbf{r_{11} }}\\
        \BAmulticolumn{2}{c}{ \mathbf{r_{12}} }\\
        \BAmulticolumn{2}{c}{ \vdots }            \\
        \BAmulticolumn{2}{c}{ \mathbf{r_{1k_2}} }\\
    \end{block}
  \end{blockarray}\right.\\
  \end{aligned} \right.
$
\normalsize
is a $k_2\times k_h$ Boolean matrix, while
$\mathbf{T}:=$
\scriptsize
$
\left. \begin{aligned}
&\left.
        \begin{blockarray}{cc}
            \begin{block}{(cc)}
        \BAmulticolumn{2}{c}{ \mathbf{r_{21}-r_{11} }}\\
        \BAmulticolumn{2}{c}{ \mathbf{r_{31}-r_{11}} }\\
        \BAmulticolumn{2}{c}{ \vdots }            \\
        \BAmulticolumn{2}{c}{ \mathbf{r_{k_11}-r_{11}} }\\
    \end{block}
  \end{blockarray}\right.\\
  \end{aligned} \right.
$
\normalsize
is a ($-1,0,1$)-Matrix with dimensions $(k_1-1)\times k_h$. Consequently, when the interior graph $\mathcal{H}$ of $\mathcal{G}$ is connected, the corresponding measurement matrix $\mathbf{R}$ can be linearly transformed to
\scriptsize
$
\left. \begin{aligned}
\left.
        \begin{blockarray}{ccccccccccc}
            \BAmulticolumn{5}{c}{\multirow{1}{*}{$W_{m_1a_1}\cdots W_{m_1a_{k_1}}$}}&\BAmulticolumn{4}{c}{\multirow{1}{*}{$W_{b_1m_2}\cdots W_{b_{k_2}m_2}$}}&\BAmulticolumn{2}{c}{\multirow{1}{*}{$W_{l_1}\cdots W_{l_{k_h}}$ }}\\
            \begin{block}{(ccccc|cccc|cc)}
         1 & && & & 1 &   &  &  &  \BAmulticolumn{2}{c}{\multirow{5}{*}{\Large $\mathbf{B}$}}  \\
         1 & && & &  & 1  &  &  &    &    \\
         \vdots &  & & & & &   & \ddots &  &    &  \\
         1 & & & &   & &   &  & 1 &    &    \\
         \cline{1-11}
         -1&1&  & & &   &   &  &  &   \BAmulticolumn{2}{c}{\multirow{5}{*}{\Large $\mathbf{T}$}}     \\
         -1&& 1 & & &  &    &  &  &    &  \\
         \vdots&& &\ddots & & &   &   &  &    &   \\
         -1&&  & & 1  & &   &  &   &    &   \\
         \cline{1-11}
         &  & &   &  &  & & & & \BAmulticolumn{2}{c}{\multirow{4}{*}{\Large $\mathbf{L}$}}\\
         &  &&    &  &  &  &&&  &            \\
        &  &&    &  &  &  &&&  &           \\
        & & &    &  &  &  &&&  &            \\
    \end{block}
  \end{blockarray}\right.\\
  \end{aligned} \right..
$
\normalsize
\hfill$\blacksquare$

\subsection{Proof of Proposition \ref{lemma:3vertexConnected}}
\begin{figure}[tb]
\centering
\includegraphics[width=3.4in]{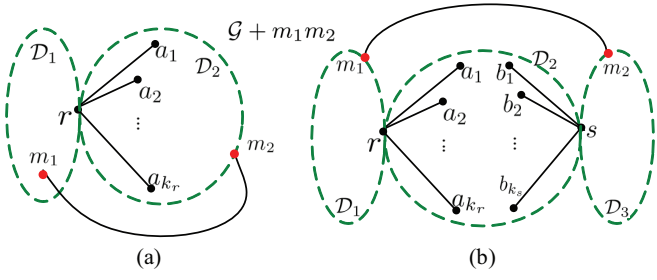}
\caption{$\mathcal{G}+m_1m_2$ is 3-vertex-connected.} \label{Fig:GreaterThan2Connected}
\end{figure}

Suppose interior graph $\mathcal{H}$ of $\mathcal{G}$ is identifiable and $\mathcal{G}+m_1m_2$ is not 3-vertex-connected, then the \emph{connectivity}\footnote{The greatest integer $k$ such that $\mathcal{G}$ is $k$-vertex-connected is the \emph{connectivity} of $\mathcal{G}$.} of $\mathcal{G}+m_1m_2$ is 1 or 2.

\emph{1)} Consider the case that the connectivity of $\mathcal{G}+m_1m_2$ is 1 and node $r\in V(\mathcal{G}+m_1m_2)$ is deleted. According to the assumption of $\mathcal{H}$ being a connected graph, $\mathcal{G}$ must be 1-vertex-connected. First, if $r$ is a monitoring node, the resulting $\mathcal{G}+m_1m_2-r=\mathcal{G}-r$ is not disconnected since $\mathcal{H}$ is connected. Second, we assume $\mathcal{G}$ is separated into two components, denoted by $\mathcal{D}_1$ and $\mathcal{D}_2$, after deleting $r$ ($r\in V(\mathcal{H})$). If each of $\mathcal{D}_1$ and $\mathcal{D}_2$ contains a monitoring node, then link $m_1m_2$ connects $\mathcal{D}_1$ and $\mathcal{D}_2$ again. If one of $\mathcal{D}_1$ and $\mathcal{D}_2$, say $\mathcal{D}_1$, does not have monitoring nodes in it, then all $m_1\rightarrow m_2$ paths employing links in $\mathcal{D}_1$ must both enter and leave $\mathcal{D}_1$ at $r$, thus forming a cycle in this path construction. Therefore, the connectivity of graph $\mathcal{G}+m_1m_2$ must be greater than 1.

\emph{2)} Now we assume the connectivity of $\mathcal{G}+m_1m_2$ is 2. First, Consider deleting one monitoring node and a non-monitoring node (displayed in Fig. \ref{Fig:GreaterThan2Connected}-a). If deleting $r$ ($r\in V(\mathcal{H})$) and $m_2$ results in $\mathcal{D}_2$ being separated, then $\mathcal{D}_1$ and $\mathcal{D}_2$ only have one common node, denoted by $r$. In this case, $m_1$ must be in $\mathcal{D}_1$; otherwise, a cycle is formed when using links in $\mathcal{D}_1$ to construct $m_1\rightarrow m_2$ paths. However, even if all link metrics in $\mathcal{D}_1$ have been identified, $(W_{ra_i})^{k_r}_{i=1}$ ($a_i\neq m_2$ and $k_r\geq 2$, since $\mathcal{G}$ is proved to 2-edge-connected in \cite{MaNetworkTomography12} when its interior graph is identifiable) are uncomputable, according to Corollary III.2 in \cite{MaNetworkTomography12}. This contradicts the claim that $\mathcal{H}$ is identifiable. Second, if $m_1$ and $m_2$ are deleted, the remaining graph is $\mathcal{H}$, which is connected, according to the assumption that the interior graph of $\mathcal{G}$ is connected. Third, now we consider deleting $r$ and $s$ ($r,s\in V(\mathcal{H})$, $r\neq s\neq m_1 \neq m_2$). If $\mathcal{G}$ is separated and each component has a monitoring node, then $m_1m_2$ can connect these two components again. If the separated component does not have monitoring nodes in it (such as $\mathcal{D}_2$ in Fig. \ref{Fig:GreaterThan2Connected}-b), then according to Corollary III.2 in \cite{MaNetworkTomography12}, $(W_{ra_i})^{k_r}_{i=1}$ and $(W_{sb_i})^{k_s}_{i=1}$ are uncomputable even if all link metrics in $\mathcal{D}_1$ and $\mathcal{D}_3$ have been identified. This also contradicts the claim that $\mathcal{H}$ is identifiable.

Thus, the connectivity of $\mathcal{G}+m_1m_2$ is greater than 2. Therefore, $\mathcal{G}+m_1m_2$ is 3-vertex-connected when its interior graph is identifiable.
\hfill$\blacksquare$

\subsection{Proof of Proposition \ref{Lemma-3-connected}}
\emph{Necessary part}.

\emph{1)} If $\mathcal{G}$ is separated by deleting 2 non-monitoring nodes, then each component must have a monitoring node; otherwise, $\mathcal{G}+m_1m_2$ is 2-vertex-connected.

\emph{2)} If one of these deleted 2 nodes is a monitoring node, say $m_1$, then $m_1m_2$ is deleted as well. Deleting any other node except $m_2$ will not result in the separation of G. If separated, $\mathcal{G}+m_1m_2$ is 2-vertex-connected.

\emph{3)} If $m_1$ and $m_2$ are deleted, we can obtain sub-graph $\mathcal{H}$, which is connected according to the assumption.

\emph{Sufficient part}.

\emph{1)} When 2 nodes are deleted in $\mathcal{G}$, if it remains connected, then $\mathcal{G}$ is 3-vertex-connected, so is $\mathcal{G}+m_1m_2$.

\emph{2)} If $\mathcal{G}$ is separated after deleting two nodes and each separated component has a monitoring node, then these components are connected again by link $m_1m_2$. Therefore, $\mathcal{G}+m_1m_2$ is 3-vertex-connected.
\hfill$\blacksquare$

\subsection{Proof of Lemma \ref{Lemma-twoCycles}}
\emph{1)} For $vw\in L(\mathcal{H})$, an H-path\footnote{$\mathcal{P}$ ($||\mathcal{P}||\geq 1$) is an \emph{H-path} of graph $\mathcal{H}$ if $\mathcal{P}$ meets $\mathcal{H}$ exactly in its end nodes.} $\mathcal{P}_1$ from $v$ to $w$ can be discovered for a 2-vertex-connected graph, according to Proposition 3.1.3 \cite{GraphTheory2005} ($\mathcal{G}$ is a 2-vertex-connected graph, since $\mathcal{G}+m_1m_2$ is 3-vertex-connected), then a cycle $\mathcal{C}'_1=\mathcal{P}_1+vw$ is formed. If $xy\in L(\mathcal{G})$ with $x,y\in V(\mathcal{C}'_1)$ and $xy \notin L(\mathcal{C}'_1)$, then use $xy$ to replace $\mathcal{P}_{\mathcal{C}'_1}(x,y)$ recursively, i.e., $\mathcal{C}'_1=\mathcal{C}'_1\setminus \overset{\circ}{\mathcal{P}}_{\mathcal{C}'_1}(x,y)+xy$, until no such $xy$ exists, where $\mathcal{P}_{\mathcal{C}'_1}(x,y)$ is the path from $x$ to $y$ in $\mathcal{C}'_1$ with $vw \notin L(\mathcal{P}_{\mathcal{C}'_1}(x,y))$. Finally, $\mathcal{C}''_1=\mathcal{C}'_1$ is an induced cycle. If $\mathcal{C}''_1$ is not a face, then there exists separated component $\mathcal{D}$ ($\mathcal{D}\cap \{m_1, m_2\}=\emptyset$, when $\mathcal{C}''_1$ is deleted), in which \emph{all} paths from $V(\mathcal{D})$ to $m_1$ or $m_2$ must have three or more (since $\mathcal{G}+m_1m_2$ is 3-vertex-connected) common nodes with $\mathcal{C}''_1$. For any $v_1\in V(\mathcal{D})$, degree\footnote{The \emph{degree} of node $v$ is the number of links incident with $v$, denoted by $d(v)$.} $d(v_1)\geq 2$, since $\mathcal{G}$ satisfies Condition \textcircled{\small 1}\normalsize. Furthermore, there must be an inner path $\mathcal{P}_{in}(x_1,x_2)$ incident with $x_1$ and $x_2$ ($x_1,x_2\in V(\mathcal{C}''_1)$) and an interior node $v_2\in V(\mathcal{D})$ with $v_2\in V\big(\mathcal{P}_{in}(x_1,x_2)\big)$. Using $\mathcal{P}_{in}(x_1,x_2)$ to replace $\mathcal{P}_{\mathcal{C}''_1}(x_1,x_2)$ ($vw \notin L(\mathcal{P}_{\mathcal{C}''_1}(x_1,x_2))$) in $\mathcal{C}''_1$ recursively, i.e., $\mathcal{C}''_1=\mathcal{C}''_1\setminus\overset{\circ}{\mathcal{P}}_{\mathcal{C}''_1}(x_1,x_2)\bigcup{\mathcal{P}}_{in}(x_1,x_2)$, until no such $\mathcal{P}_{in}(x_1,x_2)$ exists. As a result, a face ${\mathcal{C}_1}={\mathcal{C}''_1}$ can be discovered in $\mathcal{G}$.

\emph{2)} Suppose $\mathcal{C}_2$ satisfying (b) in Lemma \ref{Lemma-twoCycles} cannot be discovered, then $\mathcal{C}_2$ must share some common links with $\mathcal{C}_1\setminus \{vw\}$. Let $rs$ be one of these common links, then if $vw$ is deleted, all possible paths connecting $v$ and $w$ must traverse link $rs$. In this case, $rs$ becomes a bridge, contradicting Condition \textcircled{\small 1}\normalsize.
\begin{figure}[tb]
\centering
\includegraphics[width=1.9in]{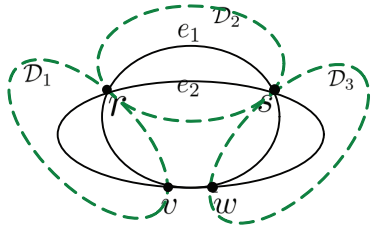}
\caption{Two cycles with two common nodes.} \label{Fig:LemmaTwoCycles}
\end{figure}

\emph{3)} Suppose there are always two common nodes no matter what strategy is used to select the two cycles. Let $r,s \in V(\mathcal{C}_1)\cap V(\mathcal{C}_2)\setminus \{v,m\}$, the two paths connecting $r$ and $s$ in $\mathcal{D}_2$ are $r\underline{e_1}s$ and $r\underline{e_2}s$ (shown in Fig. \ref{Fig:LemmaTwoCycles}). It has been proved that $r\underline{e_1}s$ and $r\underline{e_2}s$ cannot have common links, thus $\overset{\circ}{r}\underline{e_1}\overset{\circ}{s} \neq \overset{\circ}{r}\underline{e_2}\overset{\circ}{s}$. If $vw$ is deleted, any $v\rightarrow w$ paths must first traverse $r$ and then traverse $s$. Therefore, $\mathcal{G}$ is composed of three components ($\mathcal{D}_1$, $\mathcal{D}_2$ and $\mathcal{D}_3$) and link $vw$ with $V(\mathcal{D}_1 \cap \mathcal{D}_2)=\{r\}$, $V(\mathcal{D}_2 \cap \mathcal{D}_3)=\{s\}$ and $V(\mathcal{D}_1 \cap \mathcal{D}_3)=\emptyset$. For $\mathcal{D}_2$, $|V(r\underline{e_1}s\cup r\underline{e_2}s)\setminus \{r,s\}|\geq 1$, since $\overset{\circ}{r}\underline{e_1}\overset{\circ}{s} \neq \overset{\circ}{r}\underline{e_2}\overset{\circ}{s}$ and $||(\mathcal{C}_1-vw)\cap (\mathcal{C}_2-vw)||=0$. Similarly, $|\mathcal{D}_1-r-v|\geq 1$ and $|\mathcal{D}_3-s-w|\geq 1$. Only two of these three components can have $m_1$ or $m_2$. Thus, the third component without monitoring nodes is separated when the two common nodes with adjacent components are deleted, contradicting Lemma \ref{Lemma-3-connected}.

\emph{4)} $\mathcal{G}$ is connected; therefore, (d) in Lemma \ref{Lemma-twoCycles} is true.
\begin{figure}[tb]
\centering
\includegraphics[width=3.5in]{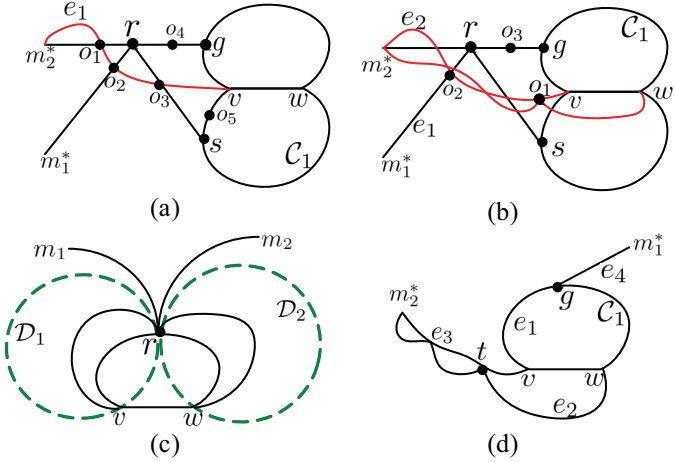}
\caption{Construction of two cycles.} \label{Fig:LemmaTwoPathsConstruction}
\end{figure}

\emph{5)} (i). If all $\mathcal{P}_1$ ($m^*_1\in V(\mathcal{P}_1)$) must traverse $m^*_2$, then $m^*_2$ is a cutvertex\footnote{A vertex which separates two other vertices in the same graph is a \emph{cutvertex}.} in $\mathcal{G}$, contradicting Lemma \ref{Lemma-3-connected}.

(ii). Let $\mathcal{C}_1=v\underline{s}w+vw$. For any $\mathcal{P}_1$ and $\mathcal{P}_2$, if they must have a common node, say $r$ (see Fig. \ref{Fig:LemmaTwoPathsConstruction}-a), then $r$ cannot be a cutvertex, because $\mathcal{G}$ is 2-vertex-connected. Therefore, there must be another path employing $v$ or $w$, say $m^*_2\underline{o_1\cdots o_5v}g$ ($r\notin V( m^*_2\underline{o_1\cdots o_5v}a)$), to connect $m^*_2$ and $g$. $m^*_2\underline{o_1\cdots o_5v}g$ might have common nodes ($o_1,\cdots, o_5$) with other paths. However, if $o_4$ or $o_5$ is the common node, then $\mathcal{P}_1$ and $\mathcal{P}_2$ do not need to traverse $r$ to connect the two cycles. If $m^*_2\underline{o_1\cdots o_5v}a$ must traverse $(o_i)^{3}_{i=1}$, then $m^*_2$ cannot connect to the two cycles when $r$ and $o_i$ are deleted, contradicting Lemma \ref{Lemma-3-connected}. Thus, an $m^*_2\underline{o_1}v$ which does not have unavoidable common nodes $o_1$, $o_2$ and $o_3$ can be constructed. Therefore, $\mathcal{C}_2$ can be reselected, i.e., $\mathcal{C}_2=v\underline{o_1ro_4g}w+vw$ with $\mathcal{P}_2=m^*_2\underline{e_1}o_1$ and $\mathcal{P}_1=m^*_1\underline{r}s$.

(iii). Let $\mathcal{C}_1=v\underline{g}w+vw$ (see Fig. \ref{Fig:LemmaTwoPathsConstruction}-b). Suppose $m^*_2$ can make use of both $v$ and $w$, say $m^*_2\underline{o_1}v$ and $m^*_2\underline{o_1}w$, to connect nodes on $\overset{\circ}{v}\underline{g}\overset{\circ}{w}$. We have $r,o_3\notin V(m^*_2\underline{o_1}v\cup m^*_2\underline{o_1}w)$, since $\mathcal{P}_1$ and $\mathcal{P}_2$ must traverse $r$ to connect $\overset{\circ}{v}\underline{g}\overset{\circ}{w}$ when $v$ and $w$ are not used. In addition, it is impossible that $m^*_2\underline{o_1}v$ and $m^*_2\underline{o_1}w$ must have an unavoidable common node, say $o_2$, with $m^*_1\underline{e_1}r$; otherwise, $m^*_1$ cannot connect to $g$ when $o_2$ and $r$ are deleted. Thus, $m^*_2\underline{o_1}v$ and $m^*_2\underline{o_1}w$ which do not have common node $o_2$ can be discovered. Then we reselect $\mathcal{C}_2$, i.e., $\mathcal{C}_2=v\underline{o_1}w+vw$ with $\mathcal{P}_2=m^*_2\underline{e_2}o_1$ and $\mathcal{P}_1=m^*_1\underline{e_1ro_3}g$.

(iv). According to (ii) and (iii), $\mathcal{P}_1$, $\mathcal{P}_2$ and $\mathcal{C}_2$ can be discovered to make sure $\mathcal{P}_1$ and $\mathcal{P}_2$ do not have common node $r$. However, if $g$ and $s$ in Fig. \ref{Fig:LemmaTwoPathsConstruction}-b are the same node (see Fig. \ref{Fig:LemmaTwoPathsConstruction}-c), we can also prove it is impossible. In this case, $V(\mathcal{D}_1)\cap V(\mathcal{D}_2)=\{r\}$ with $m_1,m_2\notin V(\mathcal{D}_1)$ and $m_1,m_2\notin V(\mathcal{D}_2)$. For the two cycles, we have $|\mathcal{D}_1|\geq 1$ and $|\mathcal{D}_2|\geq 1$ (since $vw$ is the only common link between $\mathcal{C}_1$ and $\mathcal{C}_2$); therefore, nodes in $\mathcal{D}_1$ ($\mathcal{D}_2$) without monitoring nodes are separated when $r$ and $v$ ($w$) are deleted, contradicting Lemma \ref{Lemma-3-connected}.

(v). Therefore, $\mathcal{P}_1$ and $\mathcal{P}_2$ without common nodes can be discovered. Accordingly, it is obvious that $\mathcal{P}_1$ and $\mathcal{P}_2$ do not have common links, since a common link means two common nodes (end nodes of this link) between $\mathcal{P}_1$ and $\mathcal{P}_2$. Consequently, $\mathcal{P}_1$ and $\mathcal{P}_2$ with $\mathcal{P}_1 \cap \mathcal{P}_2=\emptyset$ can be discovered.

\emph{6)} If $L(\mathcal{P})\cap L(\mathcal{C}-v-w)\neq\emptyset$, simply use the first common node as the end node of $\mathcal{P}$.

\emph{7)} We first consider $\mathcal{P}_1$. In $\mathcal{G}-m^*_2$, if $\mathcal{P}_1$ must traverse an end node of $vw$, say $v$, to connect $m^*_1$ and a node on $\mathcal{C}_1-v-w$, then nodes on $\mathcal{C}_1-v-w$ are disconnected to $m^*_1$ when $v$ and $m^*_2$ are deleted, contradicting Lemma \ref{Lemma-3-connected}. Thus, it is impossible that $\mathcal{P}_1$ must traverse an end node of $vw$. However, if $\mathcal{P}_1$ cannot avoid one of $v$ and $w$ to connect $m^*_1$ and $\mathcal{C}_1-v-w$, then two paths can be constructed. Let $v\underline{e_1g}w+vw$ be $\mathcal{C}_1$ (see Fig. \ref{Fig:LemmaTwoPathsConstruction}-d). The constructed two paths, connecting $m^*_2$ and $g$, are $m^*_2\underline{e_3tve_1}g$ and $m^*_2\underline{e_3te_2w}g$ with $m^*_2\underline{e_3t}v \cap \overset{\circ}{v}\underline{e_1g}\overset{\circ}{w}=\emptyset$ and $m^*_2\underline{e_3te_2}w \cap \overset{\circ}{v}\underline{e_1g}\overset{\circ}{w}=\emptyset$ (If they have intersections, $\mathcal{P}_1$ does not have to traverse $v$ and $w$ to connect to a node on $\overset{\circ}{v}\underline{e_1g}\overset{\circ}{w}$). Thus, according to Lemma \ref{Lemma-3-connected}, $g$ must have a connection to $m^*_1$, $m^*_1\underline{e_4}g$, with $m^*_1\underline{e_4}g \cap m^*_2\underline{e_3}t=\emptyset$ (if $m^*_1\underline{e_4}g \cap m^*_2\underline{e_3}t\neq\emptyset$, then $\mathcal{P}_1$ does not have to traverse $v$ and $w$ to connect to a node on $\overset{\circ}{v}\underline{e_1g}\overset{\circ}{w}$). Therefore, $\mathcal{C}_2$ can be chosen as $\mathcal{C}_2=v\underline{te_2}w+vw$ with $\mathcal{P}_2=m^*_2\underline{e_3}t$ and $\mathcal{P}_1=m^*_1\underline{e_4}g$. These two cycles and paths enable $vw$ to be a non-border-link identifiable via the method proposed in Section IV-B1 of \cite{MaNetworkTomography12}. Therefore, non-border-link $vw$ is capable of constructing two cycles and $\mathcal{P}_1$, $\mathcal{P}_2$ with $v,w\notin V(\mathcal{P}_1)$ and $v,w\notin V(\mathcal{P}_2)$. When considering $\mathcal{P}_2$, the same conclusion can be obtained.
\hfill$\blacksquare$

\subsection{Proof of Proposition \ref{Proposition:FaceOneBorderLink}-(a)}
Using the method to calculate Type 1 identifiable link (Section IV-B1 of \cite{MaNetworkTomography12}), all non-border-links in $\mathcal{H}$ can be identified. While for border-links, they can be categorized into two classes: (i) Class 1. $V(\mathcal{C}_1 \cap \mathcal{C}_2)=\{v,w\}$ and all $\mathcal{P}_1$ must have a common node with $\mathcal{C}_2-v-w$, and (ii) Class 2. $V(\mathcal{C}_1 \cap \mathcal{C}_2)=\{v,w,r\}$, where $r$ is another unavoidable common node.

Let $vw$ be the border-link in $\mathcal{H}$ and $vw\in L(\mathcal{C}_1)$. All other links on $\mathcal{C}_1$ can use the same face, because $\mathcal{C}'_2$ of other links cannot be disconnected to monitoring nodes when $\mathcal{C}_1$ is deleted.
\begin{figure}[tb]
\centering
\includegraphics[width=3.5in]{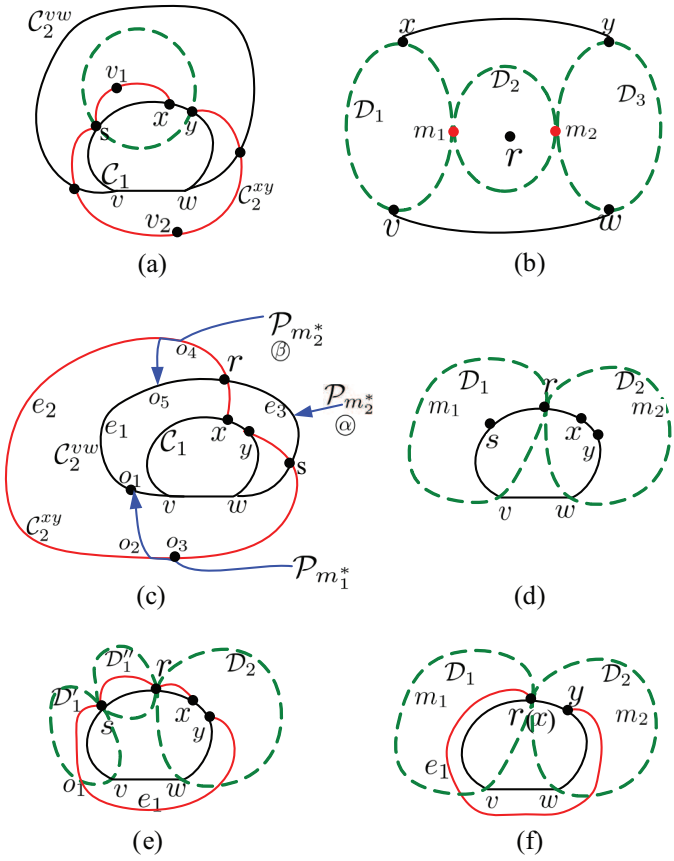}
\caption{Border link $vw$ and $xy$ cannot be in the same face.} \label{Fig:PropositionOneFaceOneBorder}
\end{figure}

\emph{1)} Let $vw$ be a border-link of Class 1.

(i). In Fig. \ref{Fig:PropositionOneFaceOneBorder}-a, suppose $xy$ is border-link of Class 2 on $\mathcal{C}_1$, then there is a common node $s$ (there is at most one common node apart from $x$ and $y$, proved in Lemma \ref{Lemma-twoCycles}) on $\mathcal{C}_1$ and $\mathcal{C}^{xy}_2$. Since $\mathcal{C}_1$ is an induced graph, there must be a node, say $v_1$, on $s\underline{v_1}x$ and a node, say $v_2$, on $s\underline{v_2}y$. For all paths connecting $v_1$ and monitoring nodes, they must traverse $s$ or $y$. Therefore, $v_1$ cannot have other connections to $\mathcal{C}_2$ via bypassing $s$ and $y$. Meanwhile, if $v_1$ has a path to one monitoring node in $\mathcal{G}\setminus \mathcal{C}^{vw}_2$, then $x$ has a path to the same monitoring node in $\mathcal{G}\setminus \mathcal{C}^{vw}_2$ as well, contradicting the assumption that $vw$ is a Class 1 border-link. Thus, when $s$ and $y$ are deleted, $v_1$ is separated from $m_1$ and $m_2$, contradicting Lemma \ref{Lemma-3-connected}. This conclusion also holds when $xy$ and $s$ have common nodes with $vw$. As the position of $s$ alters, however, the separated node might change. For instance, when $s=w$, $v_2$ is separated from $m_1$ and $m_2$ when $x$ and $w$ are deleted. Therefore, $xy$ cannot be a border-link of Class 2 on face $\mathcal{C}_1$.

(ii). Suppose there is another border-link $xy$ of Class 1 and both $\mathcal{C}^{vw}_2$ and $\mathcal{C}^{xy}_2$ must traverse $m_1$ and $m_2$. Then graph $\mathcal{G}$ can be reorganized as Fig. \ref{Fig:PropositionOneFaceOneBorder}-b, which is composed of component $\mathcal{D}_1$, $\mathcal{D}_2$, $\mathcal{D}_3$ and link $vw$, $xy$. There is at least one node, say $r$, in $\mathcal{D}_2$, because we have assumed direct link $m_1m_2$ does not exist in $\mathcal{G}$. Thus, the graph is disconnected when $m_1$ and $m_2$ are deleted, contradicting Lemma \ref{Lemma-3-connected}. Therefore, it is impossible that $\mathcal{C}^{vw}_2$ and $\mathcal{C}^{xy}_2$ must traverse both $m_1$ and $m_2$.

(iii). Since $vw$ is a Class 1 border-link, all possible $\mathcal{P}_1$ must intersect $\mathcal{C}^{vw}_2$. Thus, there exist path $\mathcal{P}_{m^*_1}:=\mathcal{P}(m^*_1, v_1)$ and $\mathcal{P}_{m^*_2}:=\mathcal{P}(m^*_2, v_2)$ with $v_1,v_2\in V(\mathcal{C}^{vw}_2)$ and $\mathcal{P}_{m^*_1}\cap \mathcal{P}_{m^*_2}=\emptyset$ (If $\mathcal{P}_{m^*_1}\cap \mathcal{P}_{m^*_2}\neq\emptyset$, the common node is a cutvertex). Suppose there is another border-link $xy$ of Class 1 on $\mathcal{C}^{}_1$ (see Fig. \ref{Fig:PropositionOneFaceOneBorder}-c). Then the associated $\mathcal{C}^{xy}_2$ ($V(\mathcal{C}^{xy}_2\cap \mathcal{C}^{}_1)=\{x,y\}$) must have two intersections (since both $vw$ and $xy$ are Class 1 border-link) with $\mathcal{C}^{vw}_2$, say $r$ and $s$ (we have proved that $r$ and $s$ cannot be both monitoring nodes in the previous step). Since $xy$ is another Class 1 border-link, if $\mathcal{P}_{m^*_1}$ connects to $\overset{\circ}{r}\underline{e_1vw}\overset{\circ}{s}$, it must have intersections with $\overset{\circ}{r}\underline{e_2}\overset{\circ}{s}$, say the intersection is $o_3$ (the number of intersections maybe greater than one, say both $o_2$ and $o_3$). In addition, we have $o_3\neq r\neq s$, since if $o_3$ overlaps with $r$ or $s$, then it means $v$ cannot connect to monitoring nodes when $r$ and $s$ are deleted, which is impossible. In Fig. \ref{Fig:PropositionOneFaceOneBorder}-c, let $o_1$ be another node, which can be equal to $v$, on $\mathcal{C}^{vw}_2$. Now we consider the locations of $\mathcal{P}_{m^*_1}$ and $\mathcal{P}_{m^*_2}$. If $\mathcal{P}_{m^*_2}$ ends at $r\underline{e_3}s$ (location \textcircled{\small $\alpha$} \normalsize in Fig. \ref{Fig:PropositionOneFaceOneBorder}-c), then $\mathcal{P}_{m^*_1}$ cannot end at $\overset{\circ}{r}\underline{e_1vw}\overset{\circ}{s}$, because $xy$ can select $x\underline{re_3s}y+xy$ as $\mathcal{C}^{xy}_2$, and then path $m^*_1\underline{o_3o_2o_1}v$ connecting $m^*_1$ and $v$ does not intersect with the newly selected $\mathcal{C}^{xy}_2$, resulting $xy$ to be a non-border-link, contradicting the assumption that $xy$ is a border-link. Therefore, $\mathcal{P}_{m^*_1}$ also ends at $\overset{\circ}{r}\underline{e_3}\overset{\circ}{s}$. In this case, however, $v$ is disconnected to monitoring nodes when $r$ and $s$ ($r$ and $s$ cannot be both monitoring nodes) are deleted, contradicting Lemma \ref{Lemma-3-connected}. Now we change the location of $\mathcal{P}_{m^*_2}$. If no $\mathcal{P}_{m^*_1}$ and $\mathcal{P}_{m^*_2}$ end at $r\underline{e_3}s$, then both $\mathcal{P}_{m^*_1}$ and $\mathcal{P}_{m^*_2}$ (location \textcircled{\small $\beta$} \normalsize in Fig. \ref{Fig:PropositionOneFaceOneBorder}-c) end at $\overset{\circ}{r}\underline{e_1vw}\overset{\circ}{s}$. In this case, $\mathcal{C}^{xy}_2$ can be reselected, i.e., $\mathcal{C}^{xy}_2=x\underline{re_3s}y+xy$ with $\mathcal{P}^{xy}_2=m^*_2\underline{o_4}r$ and $\mathcal{P}^{xy}_1=m^*_1\underline{o_3o_2o_1}v$. Thus, $xy$ with $\mathcal{P}^{xy}_1\cap \mathcal{P}^{xy}_2=\emptyset$, a Type 1 identifiable link (Section IV-B1 of \cite{MaNetworkTomography12}), is not a border-link, contradicting the assumption of $xy$ being a border-link. This conclusion also holds when $y=w$ (or $x=v$). Thus, $\mathcal{C}^{vw}_1$ cannot have another border-link of Class 1.

\emph{2)} Let $vw$ be a border-link of Class 2. For $vw$, suppose all cycles must traverse $r$, then $\mathcal{G}$ consists of component $\mathcal{D}_1$ $\mathcal{D}_2$ and link $vw$ (see Fig. \ref{Fig:PropositionOneFaceOneBorder}-d). In addition, each of $\mathcal{D}_1$ and $\mathcal{D}_2$ has a monitoring node in it; otherwise, $\mathcal{D}_1$ ($\mathcal{D}_2$) is separated from monitoring nodes when $r$ and $v$ ($w$) are deleted, contradicting Lemma \ref{Lemma-3-connected}.

(i). Suppose $xy\in L(\mathcal{D}_2)$ (see Fig. \ref{Fig:PropositionOneFaceOneBorder}-e) is a Class 2 border-link on the same face $\mathcal{C}_1$, all $\mathcal{C}^{xy}_2$ must traverse a node, say $s$, on $\mathcal{C}_1$. If $s$ is on $\overset{\circ}{v}\underline{s}\overset{\circ}{r}$, $\mathcal{D}_1$ is further split into two components ($\mathcal{D}'_1$ and $\mathcal{D}''_1$), contradicting the claim that $\mathcal{C}_1$ and $\mathcal{C}^{xy}_2$ cannot have two common nodes. Thus, $s$ cannot be on $\overset{\circ}{v}\underline{s}\overset{\circ}{r}$. If $s=r$ or $s=v$ or $s=w$, then path $o_1\underline{e_1}y$ is required. Since $vw$ is Class 2 border-link, $o_1\underline{e_1}y$ must traverse $r$ as well, resulting $\mathcal{C}^{xy}_2-xy$ containing a cycle, contradicting the basic requirement in \cite{MaNetworkTomography12}. To avoid employing cycles, $\mathcal{C}^{xy}_2$ must be in $\mathcal{D}_2$, in which case nodes, say $s$ (see Fig. \ref{Fig:PropositionOneFaceOneBorder}-d) with $s\in V(\mathcal{C}_1 \cap D_1)$, on $\mathcal{C}_1-v-w$ has a connection to $m_1$ without intersecting $\mathcal{C}^{xy}_2$. When $s\in V( \overset{\circ}{r}\underline{xy}\overset{\circ}{w})$ ($s\neq x,s \neq y$) or $x=r$ or $y=w$ or $xy\in L(\mathcal{D}_1)$, the same conclusion can be made. Thus, $xy$ cannot be a Class 2 border-link.

(ii). Suppose $xy\in L(\mathcal{D}_2)$ (see Fig. \ref{Fig:PropositionOneFaceOneBorder}-f) is a Class 1 border-link on the same face $\mathcal{C}_1$, we have $r=x$ or $r=y$, since $\mathcal{C}_1$ and $\mathcal{C}^{xy}_2$ cannot have common nodes, apart from $x$ and $y$. If $r=x$, there should be path $r\underline{e_1}y$ and $r\underline{e_1}y$ cannot have any links outside $\mathcal{D}_1$ and $\mathcal{D}_2$; therefore, $r\underline{e_1}y\subset \mathcal{D}_2$. In this case, there is a path $\mathcal{P}(m^*_1,v)$ ($r\notin V(\mathcal{P}(m^*_1,v))$. If $r$ must be on $\mathcal{P}(m^*_1,v)$, then $v$ is disconnected to monitoring nodes when $r$ and $w$ are deleted.) connecting $m^*_1$ and $v$ without intersecting $r\underline{e_1}y$, contradicting the assumption that $xy$ is a Class 1 border-link. The same conclusion can be obtained when $r=y$. Thus, $xy$ cannot be a Class 1 border-link.

Therefore, a face with a border-link cannot have another border-link.
\hfill$\blacksquare$

\subsection{Proof of Proposition \ref{Proposition:FaceOneBorderLink}-(b)}
\begin{figure}[tb]
\centering
\includegraphics[width=2in]{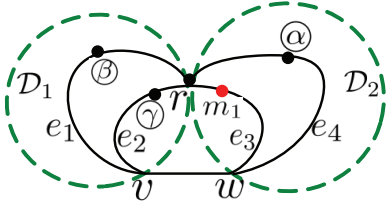}
\caption{Border-links and monitoring nodes do not in the same face.} \label{Fig:BorderLinksWithoutMonitors}
\end{figure}
If $vw$ belongs to Class 1, then all paths connecting nodes on $\mathcal{C}_1-v-w$ and monitoring nodes must intersect $\mathcal{C}_2$. Therefore, $m_1$ and $m_2$ cannot be on $\mathcal{C}_1$. If $vw$ belongs to Class 2 and all paths (besides direct link $vw$) connecting $v$ and $w$ must traverse a monitoring node, say $m_1$, then it means $r=m_1$ in Fig. \ref{Fig:PropositionOneFaceOneBorder}-d. Thus, $m_2$ is in either $\mathcal{D}_1$ or $\mathcal{D}_2$ (each component at least has two links; otherwise, the single link becomes a bridge when $vw$ is deleted). Suppose $m_2$ is in $\mathcal{D}_1$, then $\mathcal{D}_2$ is separated from monitoring nodes when $r$ ($r=m_1$) and $v$ are deleted, contradicting Lemma \ref{Lemma-3-connected}. Then obviously, it is impossible that $\mathcal{C}_1$ must traverse both $m_1$ and $m_2$. Now suppose either $m_1$ or $m_2$ must be on $\mathcal{C}_1$. Without loss of generality, let $m_1\in V(\mathcal{D}_2)$ be on $\mathcal{C}_1$ (see Fig. \ref{Fig:BorderLinksWithoutMonitors}). If $m_2$ is at location \textcircled{\small $\alpha$}\normalsize, then $\mathcal{D}_1$ is separated from monitoring nodes when $r$ and $w$ are deleted. If $m_2$ is at location \textcircled{\small $\beta$}\normalsize, then $\mathcal{C}_1=v\underline{e_2re_4}w+vw$ is reselected. If $m_2$ is at location \textcircled{\small $\gamma$}\normalsize, then $\mathcal{C}_1=v\underline{e_1re_4}w+vw$ is reselected. Therefore, for every border-link $vw\in L(\mathcal{H})$, it can discover a face without traversing $m_1$ and $m_2$. \hfill$\blacksquare$

\subsection{Proof of Proposition \ref{Proposition:FaceOneBorderLink}-(c)}
$\mathcal{P}(m^*_1,v)$ and $\mathcal{P}(m^*_2,w)$ exist, since $\mathcal{G}$ is a 2-vertex-connected graph. If $\mathcal{P}(m^*_1,v) \cap \mathcal{P}(m^*_2,w)\neq\emptyset$, let $r\in V(\mathcal{P}(m^*_1,v) \cap \mathcal{P}(m^*_2,w))$, then $v$ and $w$ cannot connect to $m^*_1$ or $m^*_2$ when $r$ is deleted, contradicting Lemma \ref{Lemma-3-connected}. Based on Proposition \ref{Proposition:FaceOneBorderLink}-(b), $m^*_1,m^*_2\notin V(\mathcal{C}_1)$. If $\mathcal{P}(m^*_1,v)\overset{\circ}{v}$ must have a common node, say $s$, with $\mathcal{C}_1$, then $m^*_1$ cannot connect to $v$ when $s$ is deleted. Thus, $\mathcal{P}(m^*_1,v)$ with $\mathcal{P}(m^*_1,v)\overset{\circ}{v} \cap \mathcal{C}_1=\emptyset$ can be found. Analogously, $\mathcal{P}(m^*_2,w)$ with $\mathcal{P}(m^*_2,w)\overset{\circ}{w} \cap \mathcal{C}_1=\emptyset$ can also be found. \hfill$\blacksquare$

\bibliographystyle{IEEEtran}
\bibliography{mybib}

\end{document}